\documentclass{article}

\usepackage{amsmath}
\usepackage{caption}
\usepackage{color}
\usepackage{epstopdf, epsfig}
\usepackage{graphicx}
\usepackage{overpic}
\usepackage{psfrag}

\def\atan{\mbox{atan}}
\def\zero{{z}}
\def\pole{{\lambda}}
\def\nTF{{N}}
\def\mTF{{M}}
\def\poleid{\pole}   
\def\sigid {\sigma}  
\def\omid  {\omega} 
\def\be{\begin{equation}}
\def\ee{\end{equation}}

\title{
\textbf{Quantifying acoustic damping using\\flame chemiluminescence}
}

\author{
\textbf{E. Boujo}$^{1}$  
\textbf{A. Denisov}$^{2}$,
\textbf{B. Schuermans}$^{3}$ \textbf{and 
N. Noiray}$^{1}$ 
}

\date{}

\begin{document}

\maketitle

\begin{center}
$^{1}$CAPS Lab., Mechanical and Process Engineering Dept., ETHZ, 8092 Z\"urich, Switzerland\\
$^{2}$Combustion Research Lab., Paul Scherrer Institute, 5232 Villigen, Switzerland\\
$^{3}$GE Power, 
5401 Baden, Switzerland
\end{center}

\graphicspath{{./}}

\begin{abstract}
Thermoacoustic instabilities in gas turbines and aeroengine combustors falls within the category of complex systems. They can be described phenomenologically using nonlinear stochastic differential equations, which constitute the grounds for output-only model-based system identification. It has been shown recently that one can extract the governing parameters of the instabilities, namely the linear growth rate and the nonlinear component of the thermoacoustic feedback, using dynamic pressure time series only. This is highly relevant for practical systems, which cannot be actively controlled due to a lack of cost-effective actuators. The thermoacoustic stability is given by the linear growth rate, which results from the combination of the acoustic damping and the coherent feedback from the flame. In this paper, it is shown that it is possible to quantify the acoustic damping of the system, and thus to separate its contribution to the linear growth rate from the one of the flame. This is achieved by post-processing in a simple way simultaneously acquired chemiluminescence and acoustic pressure data. It provides an additional approach to further unravel from observed time series the key mechanisms governing the system dynamics. This straightforward method is illustrated here using experimental data from a combustion chamber operated at several linearly stable and unstable operating conditions.
\end{abstract}

\textbf{Key words:} combustion, instability, nonlinear dynamical systems

\section{Introduction}
Thermoacoustic instabilities in modern aeroengine and gas turbine combustors is a major hurdle to overcome in order to meet ever decreasing pollutant emissions targets.
This is because the dynamic pressure load resulting from  instabilities yields high-cycle fatigue which significantly impacts the lifetime of the components, and can lead in some cases to severe damages of the combustion chamber (e.g. \cite{lieuwen_book_2012,poinsot_2016}).
Practical combustors are not equipped with active control systems for these instabilities due to the lack of cost-effective actuation technologies that would have to endure the harsh environment for several thousands of operating hours. 
The only way for  manufacturers to define an engine operating concept in order to sequentially reach the targeted operating points while avoiding 
 harmful instabilities, is to monitor the acoustic pressure or the mechanical vibrations from just a few piezosensors, accelerometers or strain gauges, and ``navigate'' through low-amplitude linearly stable regions of a multi-dimensional parameter space\footnote{Indeed, the thermoacoustic stability  depends in a non-monotonic way on several quantities like  chamber pressure,  inlet air mass flow and temperature,  hot gas temperature,  secondary air distribution mass flows,  thermal power or   fuel mass flow distribution between main and secondary injectors.}. 
In this context, it is very important to extract as much knowledge as possible about this complex system from the very limited dynamic observables. \\
In lab-scale facilities, linear growth rates $\nu$ are easily measured. 
In the linearly stable regime, the system can be forced with a harmonic excitation and $\nu$ calculated from a fit of the obtained transfer function. 
Equivalently, one can study the response of the system to an impulse forcing.
In the linearly unstable regime,  the system can be stabilised with  active control and $\nu$  deduced from the exponential growth observed after the control is turned off \cite{Poinsot92,Mejia16}.
The task is however significantly more difficult in industrial systems:  
because of the lack of cost-effective actuators, one must resort to analysing unforced time signals.
In the linearly stable regime, 
growth rates can be identified using pressure auto-correlation functions \cite{Lieuwen05}  or  pressure frequency spectra \cite{Yi08}.
Recently, it has been shown that  dynamic pressure time series contain a wealth of information, and 
can be used to develop robust output-only system identification (SI) methods even in the linearly unstable regime \cite{NoiraySchu13,NoirayDenisov16,Noiray16}, which enables the development of stability monitoring tools, the quantitative validation of linear stability prediction methods, or the design of passive damping technologies. 
\\
Turbulent reactive flows subject to thermoacoustic instabilities can be considered as complex systems with a large number of degrees of freedom, from which emerges a stochastically perturbed coherent dynamics.
This is because the ``deterministic'' limit cycle associated with the constructive thermoacoustic feedback is randomly forced by the inherent noise resulting from the highly turbulent reactive flow\footnote{This is a dynamic noise and not a measurement noise.}. 
The main macroscopic observable of the thermoacoustic coupling in  practical combustion chambers consists in a local measurement of the acoustic pressure. 
It has been shown in the aforementioned references that one can  extract the linear growth rates from the acoustic pressure signals by analysing the stochastic dynamics 
of the turbulence-driven system  around its equilibrium fixed point (in the stable regime) or  limit-cycle (in the unstable regime).
However, although this method identifies the linear growth rate $\nu$,
it does not give access to the acoustic damping $\alpha$ and flame gain  $\beta$ that  result in $\nu=(\beta-\alpha)/2$. 
Therefore, it would be particularly useful to determine these two  contributions separately, in order to gain more insight about the thermoacoustic dynamics of the considered combustion chamber. This is the  purpose of the new method presented and demonstrated experimentally in this paper. It provides individual identification of  $\alpha$ and  $\beta$ for a given thermoacoustic mode, based on the processing of acoustic pressure and flame chemiluminescence time series recorded simultaneously.

\section{Experimental measurements}
\label{sec:exp}

\begin{figure}
\psfrag{f}[t][]{$f$~(Hz)}
\psfrag{See}[r][][1][-90]{$|\mathcal{S}_{p_1 p_1}|$}
\psfrag{Sqq}[r][][1][-90]{$|\mathcal{S}_{qq}|$}
\vspace{0.3cm}
\centerline{
   \hspace{0.78cm}
   \begin{overpic}[width=9.3cm,tics=10]{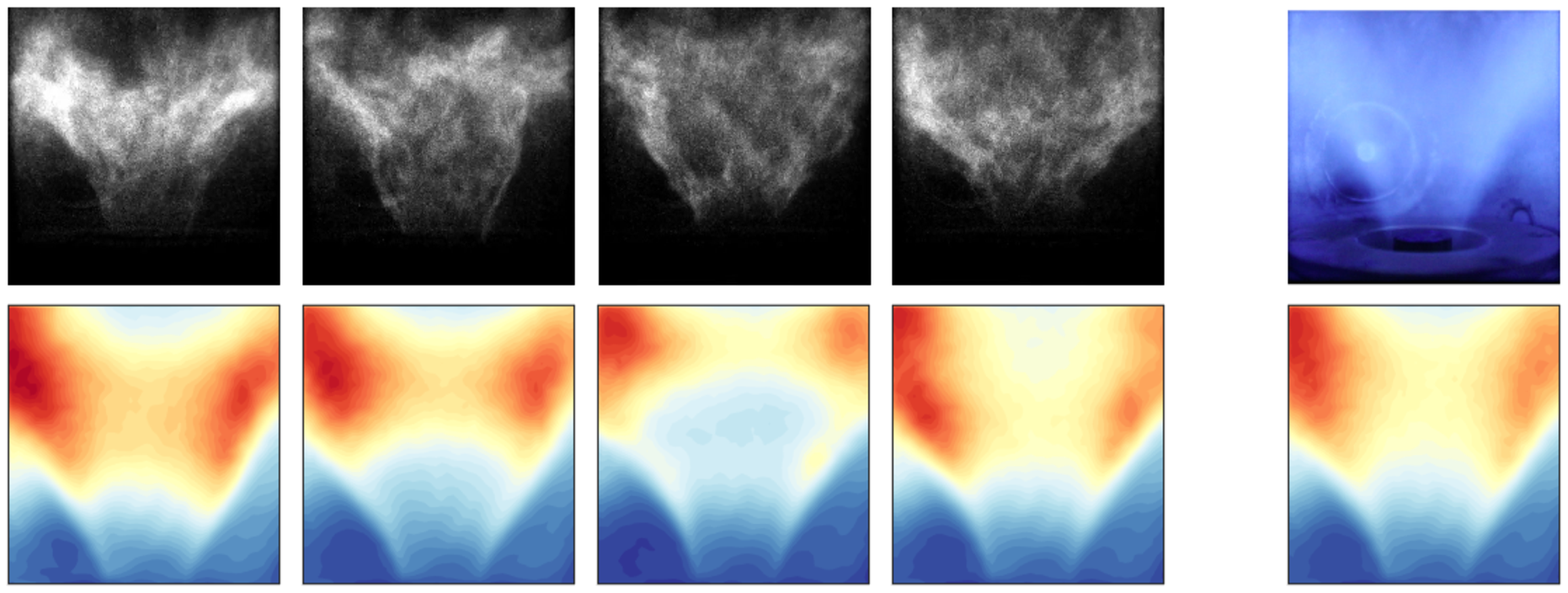}
      \put( 8,  38){  $0^\circ$}   
      \put(26,  38){ $90^\circ$}  
      \put(44.5,38){$180^\circ$} 
      \put(62.5,38){$270^\circ$} 
      \put(-4,34  ){$(a)$}   
      \put(76,34  ){$(b)$}
      \put(-4,15.2){$(c)$}   
      \put(76,15.2){$(d)$}  
   \end{overpic}
}
\vspace{0.8cm}
\centerline{
   \hspace{-0.2cm}
   \begin{overpic}[width=10cm,tics=10]{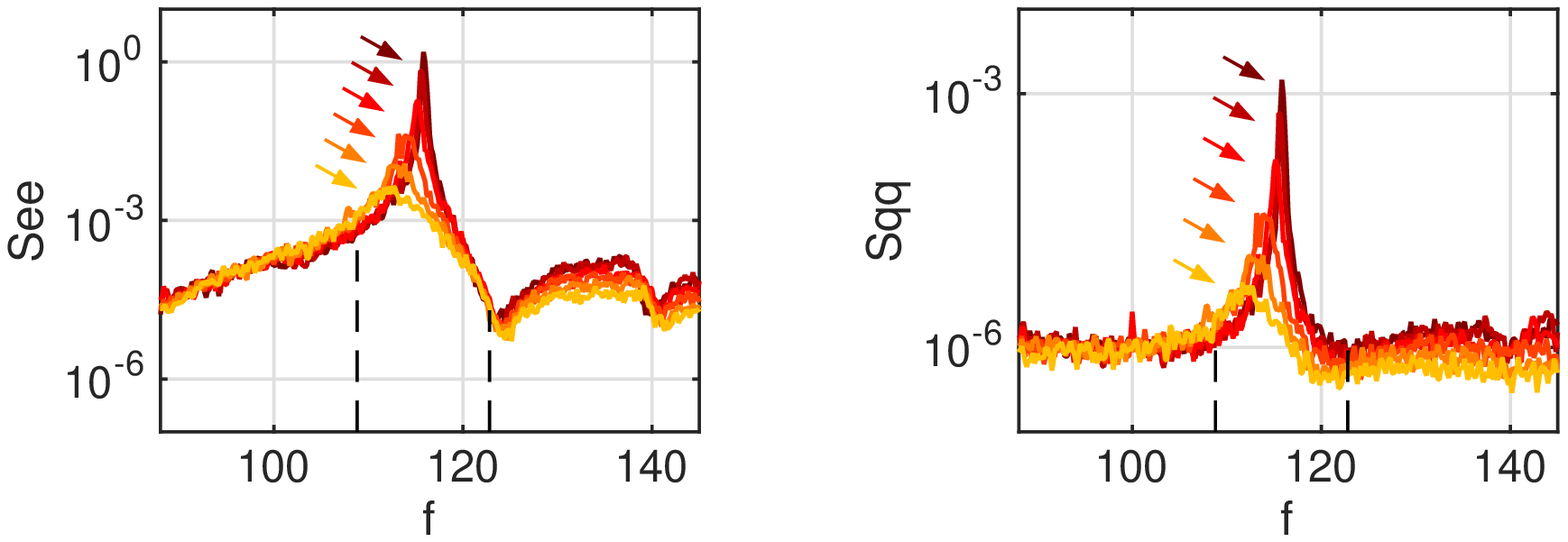}  
      \put(-5,32){$(e)$}
      \put(17,29){$f_p$}      
      \put(73,26){$f_p$}      
   \end{overpic}
}
\caption{
$(a)$ Snapshots of OH$^*$ chemiluminescence at different phases of one acoustic period,
and $(b)$~direct visualisation of the time-averaged flame.
$(c)$~Phase-averaged and $(d)$~time-averaged 
OH$^*$ chemiluminescence.
$(e)$
Power spectral density of acoustic pressure $p_1$ and heat release $q$, for different equivalence ratios $\Phi=0.521$ (light), $\ldots, 0.549$ (dark).
} 
\label{fig:flame-PSD2}
\end{figure}

\begin{figure}
\psfrag{Phi}[t][]{$\Phi$~(-)}
\psfrag{t}  [t][]{$t$~(s)}
\psfrag{Prms(eta1)}[l][][1][-90]{}
\psfrag{qrms}      [r][][1][-90]{}
\psfrag{qqq}       [t][]{$\widetilde q$~(a.u.)}
\psfrag{chem}      [t][]{$q$, $\widetilde q$~(a.u.)}
\psfrag{meanchem}  [r][][1][-90]{}
\psfrag{eta}       [t][]{$p_1$, $\widetilde p_1$~(mbar)}
\psfrag{ee}[t][]{$\widetilde  p_1$~(mbar)}
\psfrag{A} [][]{$A_1$~(mbar)}
\psfrag{Peta}[][]{$P(\widetilde  p_1)$}
\psfrag{PA}  [][]{$P(A_1)$}
\psfrag{Pq}  [][]{$P(\widetilde  q)$}
\vspace{0.5cm}
\centerline{
\hspace{-0.5cm}
   \begin{overpic}[width=14.cm,tics=10]{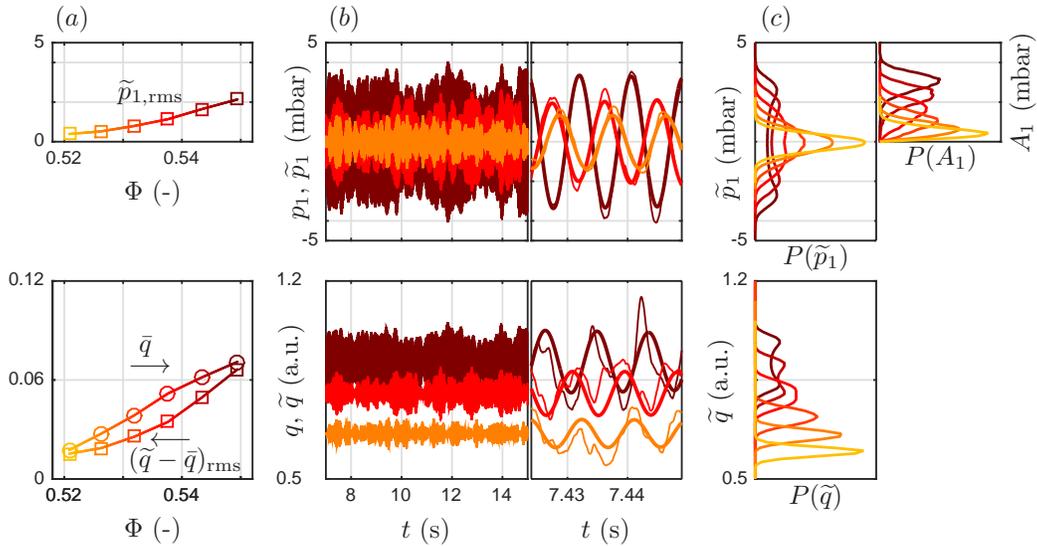}
   \put(13,41){$\widetilde p_{1,\mathrm{rms}}$}
   \put(14,15){$\longrightarrow$} 
   \put(15,17){$\bar q$}  
   \put(14,6){$(\widetilde q-\bar q)_{\mathrm{rms}}$} 
   \put(16,8){$\longleftarrow$}   
   \put( 7,48){$(a)$}
   \put(33,48){$(b)$}
   \put(74,48){$(c)$}  
   \end{overpic}
}
\caption{
$(a)$ Mean  and root mean square  of the acoustic pressure
 $\widetilde p_{1}$ 
 and heat release $\widetilde q$ filtered around the main peak frequency
vs. fuel-air equivalence ratio~$\Phi$.
$(b)$ Samples of the 180~s acoustic pressure  and heat release  signals for different equivalence ratios 
$\Phi=0.526$,  0.538,   0.549: 
original signals $p_1(t)$, $q(t)$ (thin lines) and filtered signals $\widetilde p_1(t)$, $\widetilde q(t)$ (thick lines).
$(c)$
PDF of the filtered acoustic pressure $\widetilde p_1$, of its  envelope $A_1$, and of the filtered heat release $\widetilde q$.
} 
\label{fig:signals_rms_PDFs}
\end{figure}

Experiments  are performed using a premixed methane-air, swirl-stabilised  flame anchored in an atmospheric combustion chamber  \cite{NoirayDenisov16}.
The  thermal power is  30 kW, the swirl number  approximately  0.5, the  upstream air temperature  450~K and the mean axial flow velocity downstream of the swirler  21~m/s.
The  flame is turbulent and has a typical V shape in the range of considered equivalence ratios ($0.521 \leq \Phi \leq 0.549$). 
The acoustic pressure $p_i(t)$ is measured at two locations upstream of the burner ($i$=1, 2) and two locations in the combustion chamber ($i$=3, 4)
 with water-cooled microphones (Bru\"el \& Kjaer, type 4939).
Spatially integrated line-of-sight OH$^*$ chemiluminescence intensity $I_{\mathrm{OH}^*}(t)$
 is measured with a photomultiplier equipped with a OH$^*$ filter (wavelength 310~nm).
As typically done in fully premixed configurations, one can  consider that the spatially integrated heat release rate $q(t)$ is approximately proportional to the spatially integrated chemiluminescence of OH$^*$ or CH$^*$ radicals \cite{Keller87,Docquier02,Ayoola06, Murat09, Worth13, Balusamy15, Cosic15}. 
Note that this is not true for local quantities, nor in non-fully premixed configurations \cite{Najm98, Lauer11}.
Here we use $I_{\mathrm{OH}^*}(t)$ as a measure of the integrated heat release rate.
All signals are recorded at $10$~kHz for 180~s.
The flame exhibits self-sustained oscillations illustrated  by the snapshots, time-averaged images and phase-averaged images in figure~\ref{fig:flame-PSD2}$(a-d)$.
\\
Power spectral densities (PSD)   of acoustic pressure (measured at microphone 1)
 $|\mathcal{S}_{p_1 p_1}(f)|$
and of heat release
 $|\mathcal{S}_{qq}(f)|$,
are shown in figure~\ref{fig:flame-PSD2}$(e)$ for equivalence ratios $\Phi=0.521, \ldots, 0.549$.
A clear peak is identified  that corresponds to the dominant thermoacoustic mode.
While the  frequency of this peak increases only slightly with $\Phi$, 
from $f_p=112$~Hz   
to       $116$~Hz, 
its sharpness becomes more pronounced: its height increases by about three orders of magnitude (30~dB in sound pressure level) and its quality factor (ratio of peak frequency to peak width at half-maximum of 
$|\mathcal{S}(f)|^{1/2}$) 
increases from $Q \simeq 15$ to $170$. 
In addition to this dominant peak, the acoustic PSD also contains neighbouring resonances and antiresonances, which also appear in the heat release rate PSD, albeit much weaker. 
This indicates the presence of a series of acoustic modes, among which only one interacts constructively with the flame,  leading to a thermoacoustic instability 
for the range of operating conditions considered here.
In section \ref{sec:SI}, two different SI techniques will be used to determine the governing parameters of this mode.
One technique is based on the statistics of the amplitude of the dominant mode, calculated from the acoustic signal filtered around the peak of interest (dashed lines in figure~\ref{fig:flame-PSD2}$(e)$)  \cite{NoiraySchu13}; the other, new technique is based on the thermoacoustic transfer function obtained from unfiltered pressure and chemiluminescence spectra.
It should be noted that the proximity of secondary antiresonances and  resonances  in $\mathcal{S}_{pp}(f)$ constitutes a challenging case for system identification.
\\
Figure~\ref{fig:signals_rms_PDFs} shows the effect of equivalence ratio, with low to high values of $\Phi$ indicated by light to dark colours, respectively.
Signals of acoustic pressure and heat release  exhibit a fast oscillation of period $T \simeq 9$~ms (frequency $f \simeq 110-120$~Hz) and a slower envelope modulation. Filtered signals $\widetilde p_1(t)$ and $\widetilde q(t)$ closely follow the original signals $p_1(t)$ and $q(t)$.
The amplitude of the oscillations 
$\widetilde p_{\mathrm{rms}}$ and $(\widetilde q-\bar q)_{\mathrm{rms}}$ increases substantially with $\Phi$.
The mean heat release $\bar q$ increases strongly too, consistent with observations reporting a power-law variation of $I_{\mathrm{OH}^*}$ with $\Phi$ (e.g. exponent 4.93 in \cite{Guyot09} and 5.23 in \cite{Higgins01}).
Figure~\ref{fig:signals_rms_PDFs}$(c)$ shows the probability density function (PDF) $P(\widetilde p_1)$ of the filtered acoustic pressure, which evolves continuously from unimodal (single peak centred around $\widetilde p_1=0$) for $\Phi \leq 0.532$,
 to bimodal (two symmetric peaks centred around  finite values $|\widetilde p_1|>0$) for $\Phi \geq 0.538$, typical of stable systems (noise-driven fixed point) and unstable systems (noise-driven limit cycle), respectively.
This is in  agreement with previous observations (see e.g. \cite{Lieuwen03}).
A similar transition from unimodal to bimodal PDF is observed for $P(\widetilde q)$.
The PDF $P(A_1)$ of the acoustic envelope $A_1(t)$, calculated using the Hilbert transform of $\widetilde p_1(t)$, evolves accordingly: the peak moves away from $A_1=0$, and  for  $\Phi \geq 0.538$ an inflection point appears between $A_1=0$ and the peak location.

\section{System identification}
\label{sec:SI}

\subsection{Theoretical model}
\label{sec:theo}

\begin{figure}
\psfrag{AcousticsH}[][]{\textbf{Linear acoustics} $H=\sum H_j$}
\psfrag{Acoustics}[][]{\textbf{Linear acoustics}}
\psfrag{Aceqn}[][]{$H(s)=\dfrac{s}{s^2 + \alpha s + \omega_{0}^2}  $}
\psfrag{Aceq1}[][]{$H_1(s)=\dfrac{\gamma_1 s}{s^2 + \alpha_{1} s + \omega_{0,1}^2}$}
\psfrag{Aceqj}[][]{$H_j(s)=\dfrac{\gamma_j s}{ s^2 + \alpha_{j} s + \omega_{0,j}^2}$}
\psfrag{...}[][]{$\boldsymbol{\ldots}$}
\psfrag{qtot}[][]{$q$}
\psfrag{qt}[][]{$q_t$}
\psfrag{qn}[][]{$q_t$}
\psfrag{qc}[][]{$\quad q_c$}
\psfrag{p}[][]{$p$}
\psfrag{eta}[][]{$\eta$}
\psfrag{eta1}[][]{$\eta_{1}$}
\psfrag{eta2}[][]{$\eta_{j}$}
\psfrag{FTF}[][]{\textbf{Flame response} $\quad\quad$}
\psfrag{Flame}[][]{\textbf{Flame response}}
\psfrag{FDF}[][]{FDF $F(s,|u|)$}
\psfrag{1/Z}[][]{$\dfrac{1}{Z(s)}$}
\psfrag{qc(eta)}[][]{$q_c(\eta,\dot\eta) \quad$}
\psfrag{FFF}[][]{$q_c(\eta,\dot\eta) \quad$}
\psfrag{u}[][]{$u$}
\psfrag{w}[t][]{$\omega/\omega_{0j}$}
\psfrag{G}[r][][1][-90]{$|H|$}
\psfrag{real}[b][]{}
\centerline{
  \hspace{0.cm}
  \begin{overpic}[width=13.3cm,tics=10]{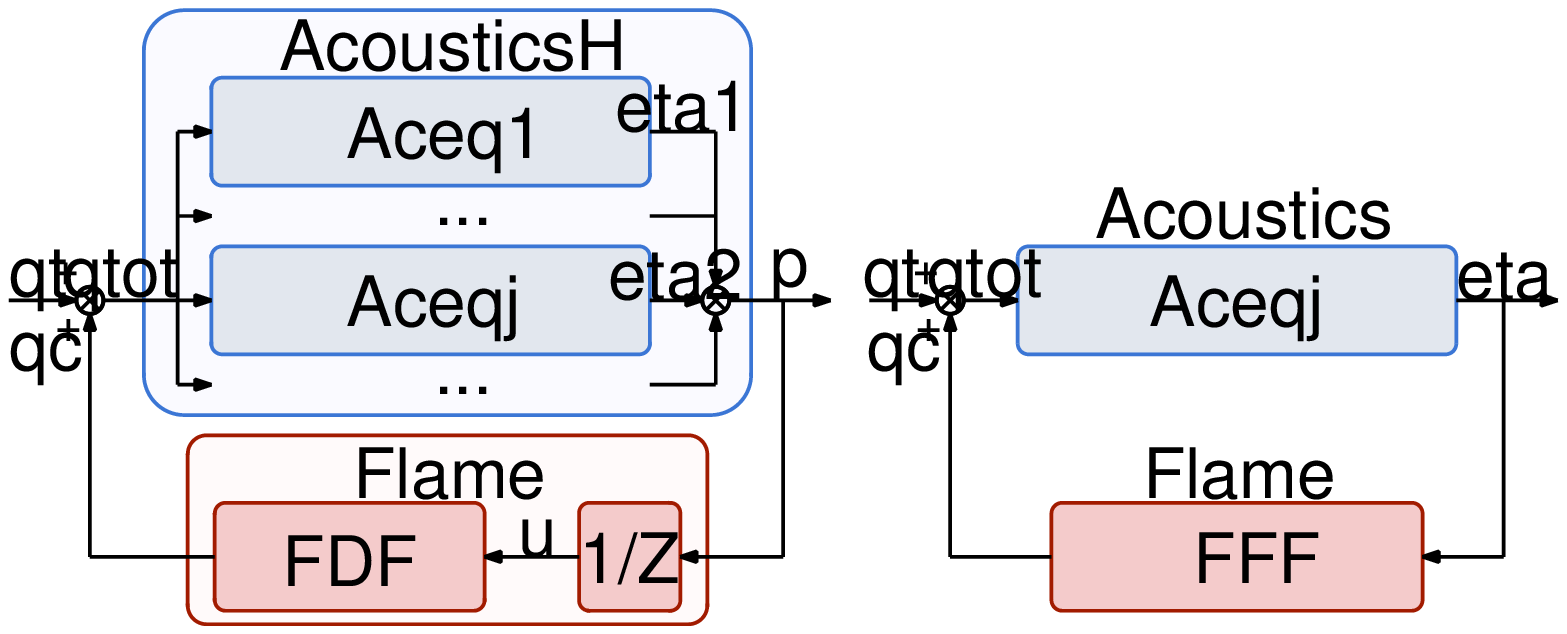} 
  \put( 1,35){$(a)$}
  \put(56,35){$(b)$}
  \end{overpic}
}
\vspace{0.5cm}
\centerline{
  \hspace{0.cm}
  \begin{overpic}[width=9cm,tics=10]{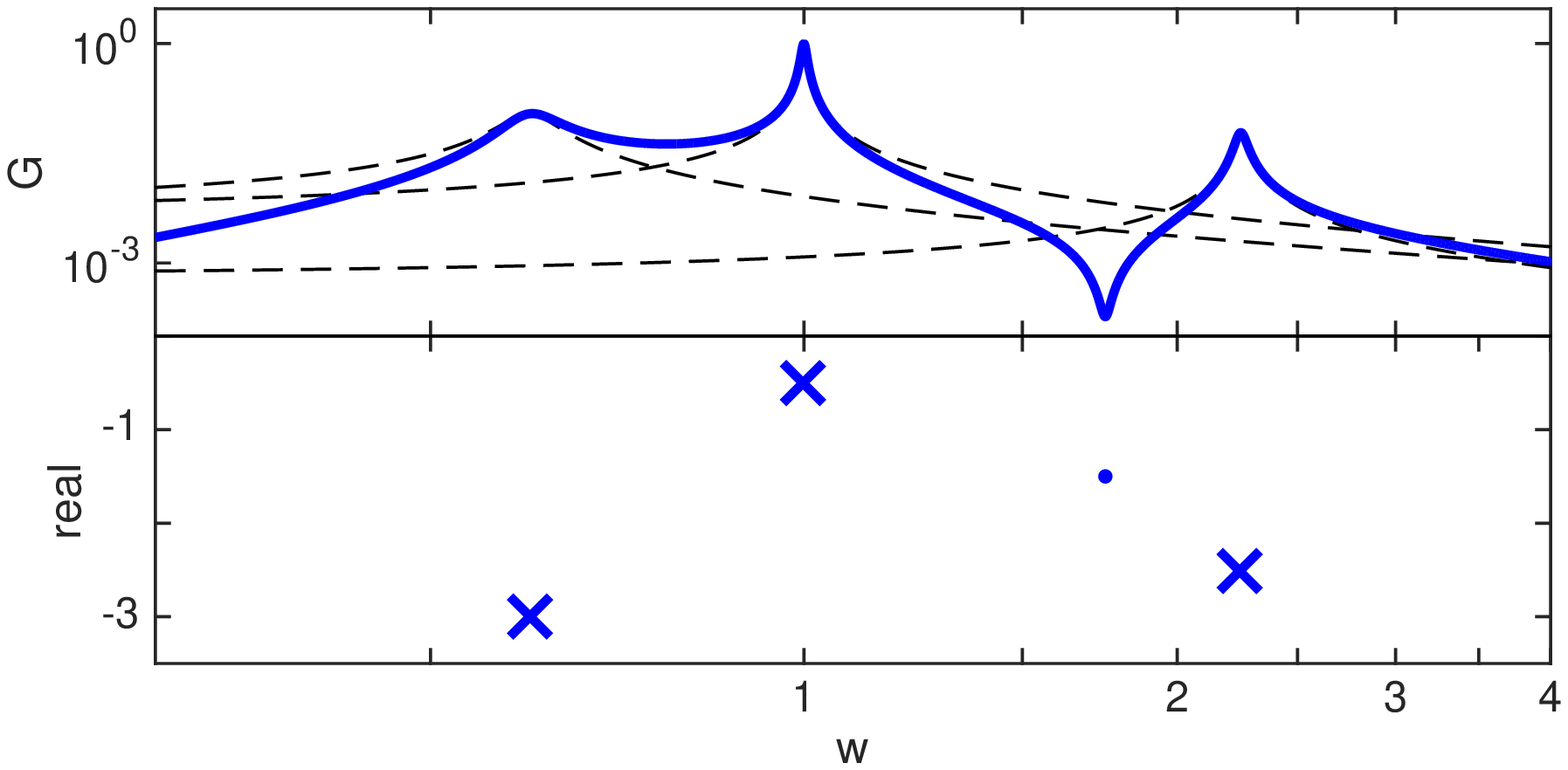}
  \put(-20, 46){$(c)$}
  \put(-20,16){$\dfrac{\Re(\lambda)}{\omega_{0j}},\dfrac{\Re(z)}{\omega_{0j}}$ ($\%$)}
  \end{overpic}
}
\caption{
$(a)$
Block diagram of the thermoacoustic system, where  $s=i\omega$ denotes the Laplace variable
and $Z$ the burner impedance.
The \textit{full acoustic transfer function} $H(s)=\sum H_j(s)$ is used in section~\ref{sec:ID2} to identify the acoustic damping  of mode $j$.
$(b)$
\textit{Single-mode approximation} used in section~\ref{sec:ID1} to identify the growth rate.
$(c)$
Sketch of the gain of a typical acoustic transfer function (---) decomposed as a sum of individual contributions (-~-).
Resonances correspond to poles~($\times$, denoted $\lambda$) and antiresonances to zeros~({\Large$\boldsymbol{\cdot}$}, denoted~$z$).
}
\label{fig:blockdiagram} 
\end{figure}

In this section we briefly recall the theoretical model that describes the thermoacoustic system; the reader is referred to \cite{NoiraySchu13}, \cite{Noiray16} and \cite{NoirayDenisov16} for more details.
Pressure is expressed in terms of acoustic modes 
\begin{equation}
p_i(t) = p(x_i,t) = \sum_j \psi_j(x_i) \eta_j(t),
\label{eq:eta0}
\end{equation}
 with 
$\psi_j(x_i)$ the spatial shape of the $j$th mode and  
$\eta_j(t)$ its time-dependent amplitude.
It follows from the wave equation that each mode satisfies a differential equation of the following form \cite{CulickAGARD,lieuwen_book_2012}:
\begin{equation}
\ddot \eta_j + \alpha_j \dot \eta_j + \omega_{0j}^2 \eta_j = \gamma_j \dot q.
\label{eq:eta1}
\end{equation}
In this damped harmonic oscillator formulation, $\omega_{0j}/2\pi$ is the natural frequency of the mode, 
the damping $\alpha_j$ is a real positive constant coming from the acoustic impedance at the boundaries  and from the volumetric damping,
and $\gamma_j \dot{q}$ is a forcing term originating from the flame heat release rate fluctuations. Specifically, $\dot{q}(t)$ is the sum of 
(i)~turbulence-driven fluctuations $q_t(t)$ induced by flow perturbations, which are characterised by spatial correlations that are much smaller than the acoustic wavelength,
and
(ii)~acoustically-driven coherent fluctuations $q_c(t)$ resulting from the nonlinear  flame response to the acoustic field.
The acoustic system can therefore be viewed as an input-output linear system driven by noise and nonlinear feedback (fig.~\ref{fig:blockdiagram}$(a)$).
The typical transfer function   in figure \ref{fig:blockdiagram}$(c)$ shows the signature of poles and zeros as resonances and antiresonances. 
In section~\ref{sec:ID2}, the \textit{full acoustic transfer function} $H(s)=\hat p(s)/\hat q(s)$ will be used  to identify the acoustic damping.
\\
In practical situations, the system dynamics at a given operating condition are often governed by a 
single thermoacoustic mode, $p_i(t) \simeq \psi_j(x_i) \eta_j(t)$, and the pressure signal is close to harmonic \cite{Culick76a,Lieuwen03}.
In this single-mode approximation, one can focus on 
the dominant frequency  
$\omega_{0}=\omega_{0j}$, and without loss of generality the coherent nonlinear forcing term 
can be expressed  at this frequency as $q_c=q_c(\eta_j,\dot\eta_j)$ 
and   conveniently expanded as a Taylor series   
\begin{align}
\gamma \dot q_c(\eta,\dot\eta) 
= \sum_{n,m} a_{n,m} \eta^n \dot\eta^m
= a_{0,1} \dot\eta + a_{1,0} \eta + \gamma \dot q_{c,nl}
\label{eq:Taylor}
\end{align}
where subscripts $j$ have been omitted.
At small amplitude, linear terms dominate.
The linear term $a_{0,1} \dot\eta$ will affect the linear stability of the oscillator; we denote its coefficient $\beta=a_{0,1}$, which  can be positive or negative depending on the convective delays involved in the response of the flame at $\omega_0$.
The linear term $a_{1,0} \eta$ will affect the oscillation frequency. We first focus on stability properties and neglect this term; it will be reintroduced in section~\ref{sec:ID2} to explain the frequency shift observed in our measurements.
Higher order terms $\dot q_{c,nl}$ describe nonlinear effects coming into play at larger amplitude, resulting for instance in saturation or bistability.
Finally (\ref{eq:eta1}) reads
\begin{equation}
 \ddot \eta -(\beta -\alpha) \dot \eta + \omega_{0}^2 \eta =
\gamma \dot q_{c,nl} +   
\gamma \dot q_t 
\label{eq:eta3}
\end{equation}
and  linear stability is determined by the sign of the growth rate $\nu=(\beta-\alpha)/2$: the system is  stable when $\nu<0$ and becomes  unstable via a Hopf bifurcation when $\nu>0$. 
In section~\ref{sec:ID1}, this \textit{single-mode approximation} will be used to identify the growth rate.

\subsection{Identification of the growth rate from pressure data}
\label{sec:ID1}

Several system identification methods have been proposed to determine the linear growth rate $\nu$ based on acoustic measurements \cite{NoiraySchu13}.
These methods rely on the stochastic nature of $q_t(t)$ which drives the system away from its deterministic equilibrium: the identification of the parameters governing the system dynamics can be done by processing data and analysing statistical quantities. \\
Non-coherent heat release rate fluctuations are well modelled by a white noise $\gamma \dot q_t(t)=\xi(t)$ of intensity $\Gamma$, since the power spectrum of
turbulence-induced heat release rate fluctuation
decays smoothly \cite{Rajaram09} and does not vary substantially in the frequency range of the sharp thermoacoustic peak. 
Then (\ref{eq:eta3}) reads 
\begin{equation}
\ddot \eta - 2\nu \dot \eta + \omega_0^2 \eta 
= 
\gamma \dot q_{c,nl} + \xi.
\end{equation}
In combustion chambers, the growth rate, the nonlinearity and the stochastic forcing are usually such that the system's oscillations are close to harmonic and conveniently described by their slowly varying envelope amplitude and phase:
\begin{equation}
A(t)=(\eta^2+(\dot\eta/\omega_0)^2)^{1/2},
\quad
\varphi(t)=-\atan(\dot\eta/\omega_0 \eta)-\omega_0 t.
\end{equation}
Note that  $A^2$ is proportional to the total acoustic energy (potential and kinetic).
Deterministic and stochastic averaging \cite{Strato2} 
yields a set of Langevin equations 
\begin{equation}
\displaystyle \dot{A} = \nu A - \frac{\kappa}{8} A^3 + \frac{\Gamma}{4 \omega_0^2 A} + \zeta
= -\frac{\mathrm{d} \mathcal{V}}{\mathrm{d} A} + \zeta, 
\quad
\displaystyle \dot{\varphi} = \frac{1}{A} \chi,
\label{eq:Langevin}
\end{equation}
where 
$\mathcal{V}(A) = -\nu A^2/2 + \kappa A^4/32 - (\Gamma/4 \omega_0^2) \ln(A)$ 
is the 
potential governing the dynamics of $A$, defined up to an additive constant,
and 
$\zeta(t)$ and $\chi(t)$ are white noises of intensity $\Gamma/2\omega_0^2$.
Terms in $A^0$ and $A^2$  are  negligible in the phase equation and are omitted here, while
the equation for the envelope amplitude is exact up to $A^4$ for any nonlinearity $\dot q_{c,nl}(\eta,\dot\eta)$ \cite{Lieuwen03,NoiraySchu13}.
In particular, 
if coherent heat release rate fluctuations were  modelled by a simple cubic nonlinearity 
$\gamma q_c(\eta,\dot\eta) = \beta \eta - (\kappa/3) \eta^3$, the acoustic pressure would be governed by 
the stochastic differential equation of a noise-driven Van der Pol oscillator \cite{NoiraySchu13}
\begin{equation}
\ddot \eta - 2\nu \dot \eta + \omega_0^2 \eta =  - \kappa \eta^2 \dot \eta + \xi,
\label{eq:eta2}
\end{equation}
however more general nonlinearities would lead at order $A^4$ to the same Langevin equations~(\ref{eq:Langevin}). 
Note that this procedure is very general: should higher-order terms be needed to describe the flame response $q_c(\eta,\dot \eta)$, they could readily be included\footnote{e.g. subcritical Hopf bifurcation, or sigmoid type saturation for supercritical bifurcations \cite{Noiray16,Gopal2016}.}; for the sake of clarity, here we illustrate the method  with the abovementioned cubic nonlinearity, valid for a range of amplitudes in the case of super-critical Hopf bifurcations. 
\\
The Langevin equation for the acoustic pressure envelope $A$ in (\ref{eq:Langevin}) is associated with a Fokker-Planck (FP) equation that describes the time evolution of the PDF of $A$ and whose long-time solution is the stationary PDF 
\begin{equation}
P(A) = \mathcal{N} \exp \left( -(4\omega_0^2/ \Gamma) \, \mathcal{V(A)} \right),
\label{eq:PDF_stat}
\end{equation}
with $\mathcal{N}$ a normalization coefficient such that $\int_0^\infty P(A) \,\mathrm{d}A = 1$ \cite{Risken84}.
This approach was followed by \cite{Lieuwen03} to describe  modifications of $P(A)$ when the system transitions from stable to unstable.
The analytical expression (\ref{eq:PDF_stat}) can be  further used to identify the growth rate $\nu$ unambiguously via a fit of the measured PDF, combined with the fit of another statistical quantity (PDF $P(\eta)$ or $P(A \dot \varphi)$, power spectrum $\mathcal{S}_{\eta\eta}(f)$ or $\mathcal{S}_{AA}(f)$,  auto-correlation function, etc.);
alternatively, one can identify the system's parameters by fitting only the coefficients of the abovementioned FP equation \cite{NoiraySchu13}.
\begin{figure}
\def \thiswidth {4.2cm}
\psfrag{PPhi}[t][]{$\Phi$~(-)}
\psfrag{Phi=0.549}[b][]{$\Phi=0.549$}
\psfrag{Phi=0.538}[b][]{$\Phi=0.538$}
\psfrag{Phi=0.526}[b][]{$\Phi=0.526$}
\psfrag{A}[t][]{$A$~(mbar)}
\psfrag{AA}[][]{$A$~(mbar)}
\psfrag{PA}[][]{$P(A)$}
\psfrag{VA}[][]{$\mathcal{V}(A)$}
\psfrag{PP2}[r][][1][-90]{}
\psfrag{PP4}[r][][1][-90]{}
\psfrag{PP6}[r][][1][-90]{$P(A)$}
\psfrag{VV2}[r][][1][-90]{}
\psfrag{VV4}[r][][1][-90]{}
\psfrag{VV6}[r][][1][-90]{$\mathcal{V}(A)$}
\centerline{ 
   \begin{overpic}[width=13.5cm,tics=10]{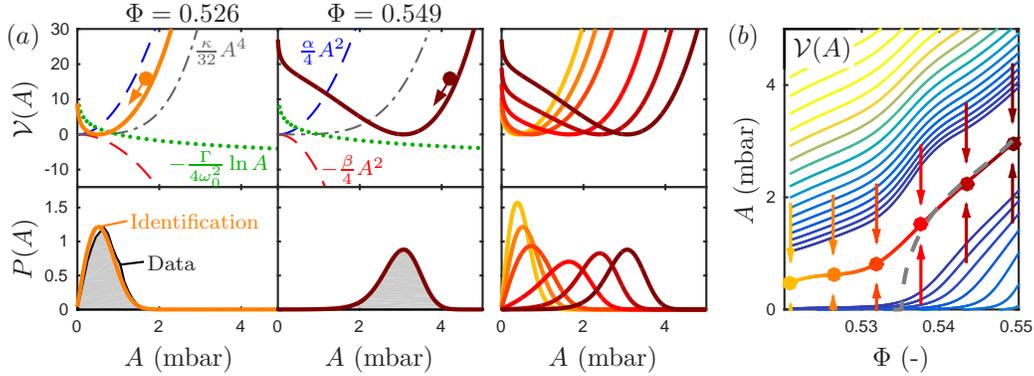} 
   \put(-1, 31){$(a)$}
   \put(70, 31){$(b)$}
   \put(10.7, 33.1){$\Phi=0.526$} 
   \put(30.6, 33.1){$\Phi=0.549$}
   \put(27.8, 29.7){\footnotesize \textcolor{blue} {$ \frac{\alpha}{4}  A^2 $}}
   \put(17.5, 29.7){\footnotesize \textcolor[rgb]{0.4,0.4,0.4}{$ \frac{\kappa}{32} A^4 $}}           
   \put(15, 18.5){\footnotesize \textcolor[rgb]{0,  0.7,0}  {$-\frac{\Gamma}{4\omega_0^2} \ln A$}}  
   \put(30, 18){\footnotesize \textcolor{red}             {$-\frac{\beta}{4}   A^2 $}}         
   \put(11.2,12.8){\footnotesize \textcolor[rgb]{1,0.5,0}{Identification}}
   \put(12.9, 8.7){\footnotesize Data}
   \put(76.5, 29.7){$\mathcal{V}(A)$}
   \end{overpic}  
}
\caption{
$(a)$
Identified potential governing the acoustic amplitude $\mathcal{V}(A)$ (solid lines) and contributions of the different terms:
$\alpha A^2/4$ from the acoustic damping and 
$-\beta A^2/4$ from the flame linear gain (dashed lines), 
$\kappa A^4/32$ from the flame nonlinearity (dash-dotted lines) and 
$-(\Gamma/4\omega_0^2) \ln A$ from turbulence-induced noise (dotted lines); 
 PDF $P(A)$ from measurement (shaded regions) and  system identification (solid lines). 
Equivalence ratio 
$\Phi=0.526$ (left panel), $\Phi=0.549$ (center),  
and $\Phi=0.521, \ldots, 0.549$ (right).
$(b)$~Identified potential valley in the acoustic amplitude - equivalence ratio plane. 
Dots: minimum of $\mathcal{V}(A)$; 
dashed line: deduced deterministic amplitude 
$A_{det}(\Phi)=\sqrt{8\nu(\Phi)/\kappa(\Phi)}$.
}  
\label{fig:V_P}
\end{figure}
Figure~\ref{fig:V_P}$(a)$ shows the PDF and potential obtained using the latter SI method and the observed acoustic pressure time series at the different operating conditions considered in this study. 
The agreement between measured and reconstructed PDFs is excellent, suggesting that the assumptions used to derive 
(\ref{eq:Langevin}), i.e. single-mode approximation and coherent/incoherent decomposition of the heat release rate,
 hold in the present situation.
As noted in section~\ref{sec:exp}, the maximum of $P(A)$ moves towards larger amplitudes as $\Phi$ increases; by definition, the potential well follows the same displacement.
The shape of $\mathcal{V}(A)$ is determined by the relative contributions of its terms.
The term $-(\Gamma/4\omega_0^2)\ln(A)$ 
 from  stochastic averaging prevents the amplitude to vanish;
conversely, the term $\kappa A^4/32$  from the flame nonlinearity ensures that the amplitude saturates to a finite value.
The term $-\nu A^2 /2=-(\beta-\alpha) A^2 /4$ is  stabilising (resp. destabilising)  when $\nu<0$ (resp. $\nu>0)$, then pushing the potential well toward small (resp. large) amplitudes.
Figure~\ref{fig:V_P}$(b)$ shows the bifurcation diagram in the $\Phi-A$ plane, with  potential  contours (thin lines), potential minimum (dots) and deduced deterministic amplitude 
$A_{det} = \sqrt{8\nu/\kappa}$ 
(dashed line).
\\
At this stage, only the growth rate $\nu=(\beta-\alpha)/2$ has been identified. The method which provides the individual contributions of the damping $\alpha$ and of the source strength $\beta$  shown in figure~\ref{fig:V_P}$(a)$ is presented in the following section.

\subsection{Identification of the acoustic damping from transfer function fitting}
\label{sec:ID2}

Here we propose a new technique that allows the individual identification of acoustic damping  $\alpha$ and flame gain $\beta$.
Unlike the  output-only  method of section~\ref{sec:ID1}, based on acoustic pressure, 
the present method is based on two sets of simultaneously acquired data:  acoustic  pressure $p(t)$ as output, and  heat release $q(t)$ as input\footnote{It is important to emphasise the fact that there is no external forcing using loudspeaker or any kind of actuation in the system, and that the input used for this single-input-single-output (SISO) system identification is the natural flame chemiluminescence recorded by the photomultiplier, assumed to be proportional to the heat release rate.}  according to the block diagram shown in  figure~\ref{fig:blockdiagram}.
To the authors' knowledge, this method has never been applied to quantify acoustic damping rates, although its principle and practical implementation are simple.
The idea is to fit  the measured acoustic transfer function 
$H(s) = \hat p(s)/\hat q(s)$ with a model of order $\nTF$
\be 
\displaystyle
\widetilde H(s) = \prod\limits_{k=1}^{\nTF} (s-\pole_k)^{-1} 
       \prod\limits_{l=1}^{\mTF} (s-\zero_l), \quad \nTF \geq \mTF,
\label{eq:Hansatz}
\ee
where $s=i\omega$ denotes the Laplace variable.
Each complex pole $\pole_k = \sigma_k+i\omega_k$
 in (\ref{eq:Hansatz}) corresponds to  a growth rate $\sigma_k$ and a frequency $\omega_k/2\pi$  (and similarly for each zero $\zero_l$).

The dominant mode $j$ governed by (\ref{eq:eta1}) is associated with the transfer function
\be 
H_j(s) = \frac{\hat \eta_j(s)}{\hat q(s)} = \frac{\gamma_j s}{s^2 + \alpha_j s + \omega_{0j}^2}
=  \frac{\gamma_j s}{(s-\pole_a)(s-\pole_a^*)},
\label{eq:acousticTF}
\ee 
where the acoustic pole is $\pole_a = -\alpha_j/2+i \omega_a$, and the reduced pulsation $\omega_a=\sqrt{\omega_{0j}^2 - \alpha_j^2/4}$ is close to $\omega_{0j}$  in general since the damping is small compared to the pulsation ($\alpha_j \ll \omega_{0j}$).
In the following subscripts $j$ are omitted.
If the transfer function fitting is successful, the least stable pole 
$\poleid=\sigid + i\omid$ of $\widetilde H(s)$ identified in the vicinity of the frequency of interest 
is expected to yield a good estimate of $\pole_a$,
and therefore the damping is obtained as  
$\alpha \simeq -2 \sigid$.

Once the damping $\alpha$ is identified from this input-output SI, 
and with the growth rate $\nu$ available from  output-only SI (section \ref{sec:ID1}), one can retrieve the flame gain as $\beta=2\nu+\alpha$. 
It is important to mention that the OH$^*$ chemiluminescence intensity recorded here using a photomultiplier equipped with a narrowband filter is not necessarily a good indicator of the heat release rate $q$. This assumption holds in the present case where the test rig is operated under fully-premixed condition, but it is  more difficult to justify the use of this method for technically premixed configurations\footnote{Usually, the modal damping does not change significantly with the operating condition and one could imagine (when  technically feasible) identifying the damping coefficient from fully premixed operating points and keeping this estimate for technically premixed conditions.}.

\begin{figure}
\psfrag{f}[t][]{$f$~(Hz)}
\psfrag{fff}[t][]{$f$~(Hz)}
\psfrag{ffff}[t][]{$f$~(Hz)}
\psfrag{TFmag}[r][][1][-90]{$|H|$}
\psfrag{TFang}[][][1][-90]{$\,\,\angle H$}
\psfrag{growthrate}[b][]{$\sigma_k, \Re(z_l)$ (rad/s)$\quad$}
\vspace{0.3cm}
\centerline{
   \hspace{0.15cm}
   \begin{overpic}[width=8.18cm,tics=10]{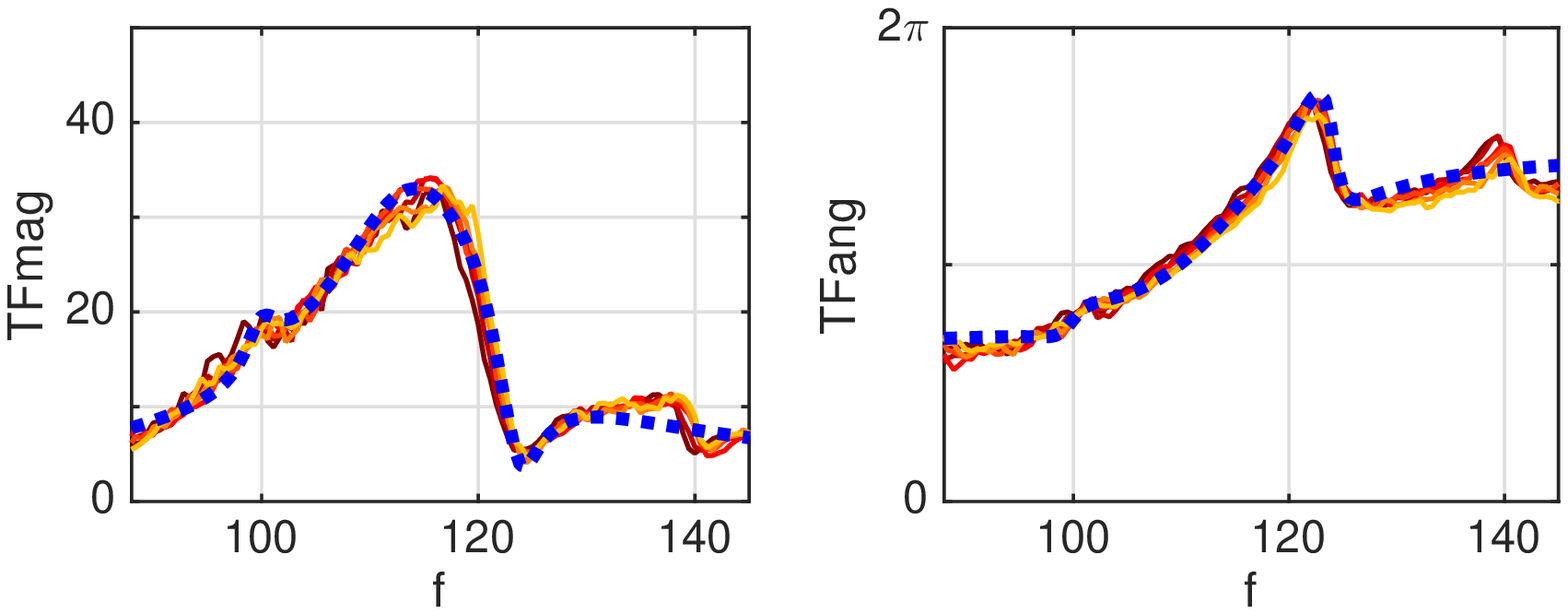}
      \put(-4,37){$(a)$}
   \end{overpic}
   \hspace{0.6cm}
   \begin{overpic}[width=3.95cm,tics=10]{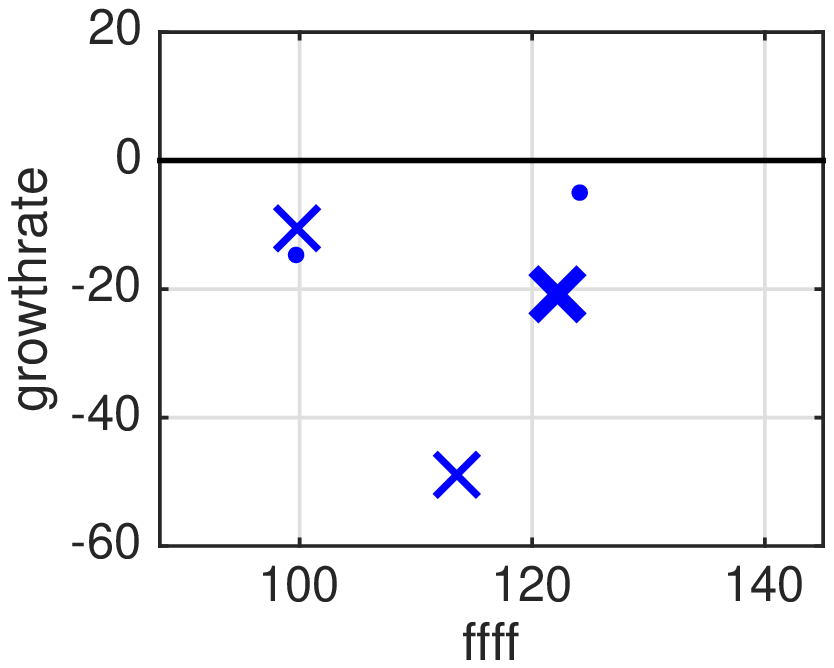}
      \put(-5,78){$(b)$} 
      \put(65,35){\small \textcolor{blue}{$(f_a,-\dfrac{\alpha}{2})$}} 
   \end{overpic}  
}
\caption{
$(a)$~Gain and phase of the acoustic transfer function (solid lines, $\Phi=0.521, \ldots, 0.549$)
and a simultaneous gain-phase fit (blue dotted line, $\Phi=0.538$, $N=6$, $\Delta f=30$~Hz).
$(b)$~Identified poles ($\times$) and zeros 
({\Large$\boldsymbol{\cdot}$}).
The bold cross indicates the dominant pole $\pole_a$ in the vicinity of the peak frequency $f_p$ (112-116 Hz, see fig.~\ref{fig:flame-PSD2}$(e)$).
} 
\label{fig:H}
\end{figure}

The  acoustic transfer function from flame heat release to  acoustic pressure at microphone 1 is calculated as the ratio of cross power spectral densities 
$H(f)=\mathcal{S}_{p_2 p_1}(f) / \mathcal{S}_{p_2 q}(f)$
(using the acoustic pressure at microphone 2 as an auxiliary signal  increases the signal-to-noise ratio, but $H$ can also be calculated as 
$\mathcal{S}_{p_1 p_1} / \mathcal{S}_{p_1 q}$ or 
$\mathcal{S}_{p_1 q}^* / \mathcal{S}_{q q}^*$).
Note that in general the transfer function  $\hat q_c/\hat p$  (linked to the flame transfer function $Z \hat q_c/\hat p = \hat q_c/\hat u$) is not related to the acoustic transfer function $H=\hat p/\hat q$  and  cannot be deduced from the present measurements since  $q$ contains both the acoustic coherent contribution $q_c$ and the incoherent turbulent contribution $q_t$ (fig. \ref{fig:blockdiagram}); only at strong resonance ($|q_c| \gg |q_t|$) can one relate 
these transfer functions 
via $\hat q_c/\hat p \simeq (\hat p/\hat q)^{-1}$.
Figure~\ref{fig:H}$(a)$ shows that the gain $|H|$ exhibits a dominant peak close to $f_p=$112-116~Hz (recall fig.~\ref{fig:flame-PSD2}$(e)$).
The overall shape  of $H(f)$ is independent of the equivalence ratio, although both heat release and acoustic pressure spectra become increasingly peaked with $\Phi$ (fig.~\ref{fig:flame-PSD2}$(e)$). 
This was to be expected: a small change of the equivalence ratio can dramatically change the flame response to acoustic perturbations, and can therefore change the system stability through $\beta$, but it should not significantly influence the acoustic damping $\alpha$, which depends mainly on the temperature, flow field and  combustion chamber geometry. 

One typical example of transfer function fitting is shown in figure~\ref{fig:H}$(a,b)$ for 
$\Phi=0.538$,
fitting order $\nTF=6$ and 
fitting interval $f_p\pm\Delta f$, $\Delta f= 30$~Hz.
Lower-order transfer functions ($\nTF=2,3$) capture the overall shape of $H(f)$  and provide a first idea of the location of the pole  associated with the dominant peak. 
The asymmetry of this peak is well captured when $\nTF=4,5$ and, as shown in fig.~\ref{fig:H}$(c)$, results from the presence of a neighbouring zero (antiresonance). 
Increasing the order further ($\nTF\geq 6$) provides finer details but does not affect the pole-zero pair in the vicinity of $f_p$ that is necessary to describe the main peak (see  supplementary materials).
It is worth emphasising that a simultaneous fit of the gain and phase of $H(f)$ is essential to identify its poles and zeros accurately. 
The robustness of the identification with respect to the fitting width $\Delta f$ is also  shown in the supplementary materials.
Thanks to the robustness of the fitting procedure with respect to $\nTF$, it is straightforward to identify the real part $\sigid$ of the dominant pole at $f_0$ 
(figure \ref{fig:H}$(b)$).
Combining results  as in figure~\ref{fig:poles_vs_n_and_phi}$(a)$ confirms that the  damping 
 $\alpha=-2\sigid$ 
of the dominant acoustic mode 
does not depend significantly on $\Phi$; 
its identified mean value is $\alpha = 37$~rad/s.
As mentioned previously,  $\alpha \ll 2\pi f_0$.

Importantly, being able to determine the acoustic damping also provides a quantification of the flame source strength $\beta=2\nu+\alpha$. 
Using the values of $\nu$ from section \ref{sec:ID1} yields the evolution of $\beta$ depicted in figure~\ref{fig:poles_vs_n_and_phi}$(a)$.
It appears that the thermoacoustic system becomes unstable as the strength of the flame source term increases and overcomes the  acoustic damping for $\Phi \simeq 0.536$.
Knowing $\alpha$ and $\beta$ individually (rather than the growth rate $\nu=(\beta-\alpha)/2$ alone) is particularly valuable  since it allows  for a better quantitative understanding of how $\nu$ varies,
either when  the acoustic properties of the system are modified (e.g. when using acoustic dampers,
minimising leakages or transferring the burner from a development test rig to the final combustion chamber),
or when the flame properties are modified (e.g. change in operating conditions such as mean flow velocity, swirl number, equivalence ratio, etc.).

\begin{figure}
\def \sigx {43}
\psfrag{phi}[t][]{$\Phi$~(-)}
\def \thistext {$\sigid=-\alpha/2$}
\psfrag{n}[t][]{$\nTF$}
\psfrag{growthrate}[][]{$\sigma$~(rad/s)}
\psfrag{gg}[r][][1][-90]{}
\psfrag{freq}[r][][1][-90]{$f$~(Hz)}
\psfrag{sigma}[][]{$\sigma$~(rad/s)}
\psfrag{alpha}[r][][1][-90]{$\alpha$~(rad/s)}
\psfrag{beta}[r][][1][-90]{$\beta$~(rad/s)}
\psfrag{ab}[][]{$\alpha, \beta$~(rad/s)}
\psfrag{aabb}[][]{$\alpha, \beta$~(rad/s)}
\psfrag{2nu}[][]{$2\nu$~(rad/s)}
\psfrag{f}[t][]{$f$~(Hz)}
\psfrag{b;2sig=-a}[b][]{growth rate~(rad/s)}
\psfrag{ab}[][]{}
\vspace{0.2cm}
\centerline{
   \begin{overpic}[height=4cm,tics=10]{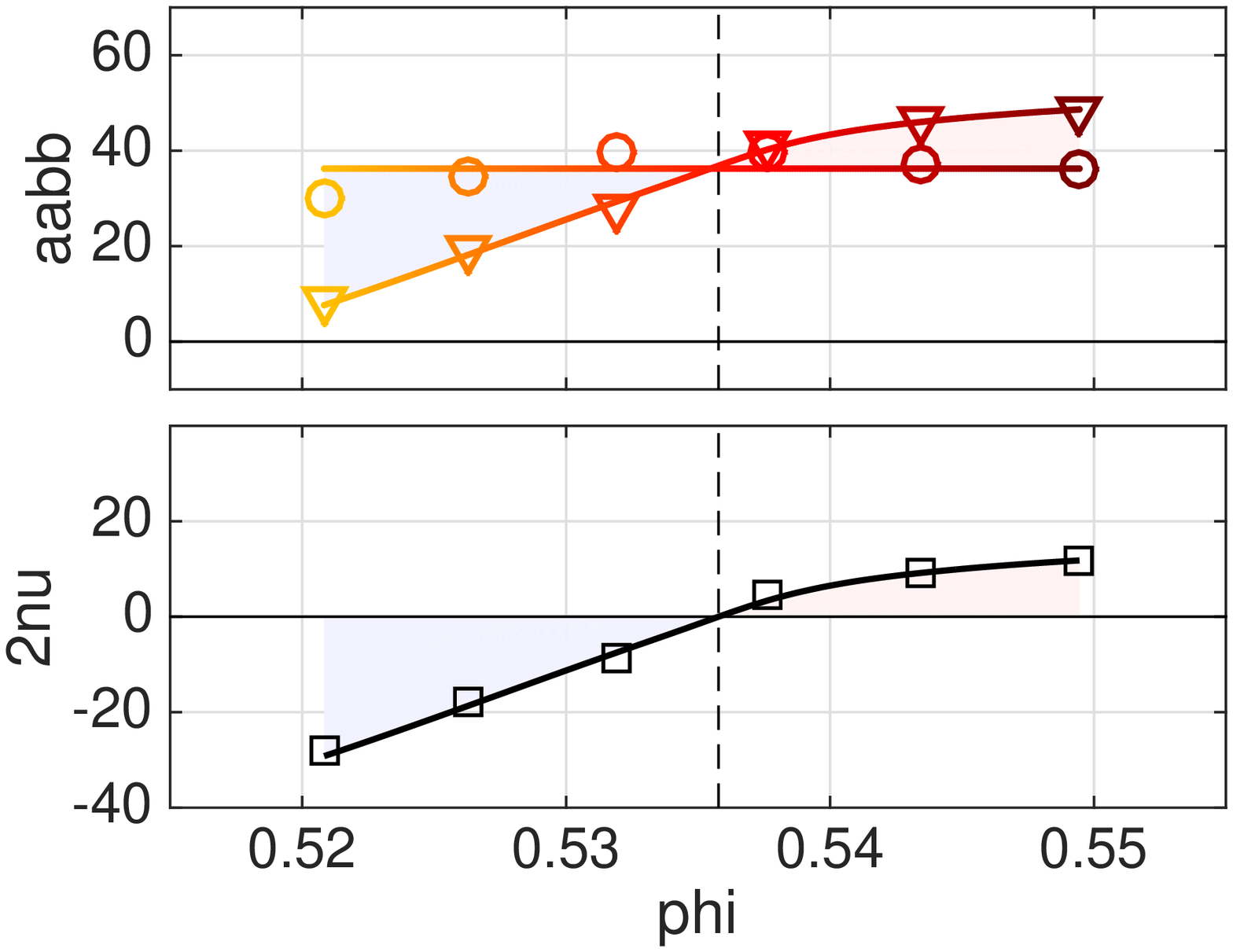}
      \put(-11,76){$(a)$}
      \put(29,66){$\alpha$}
      \put(42,52){$\beta$}
      \put(39,15){\textcolor{blue}{stable}}
      \put(65,35){\textcolor{red}{unstable}}
   \end{overpic}   
   \hspace{0.9cm}
   \begin{overpic}[height=4.1cm,tics=10]{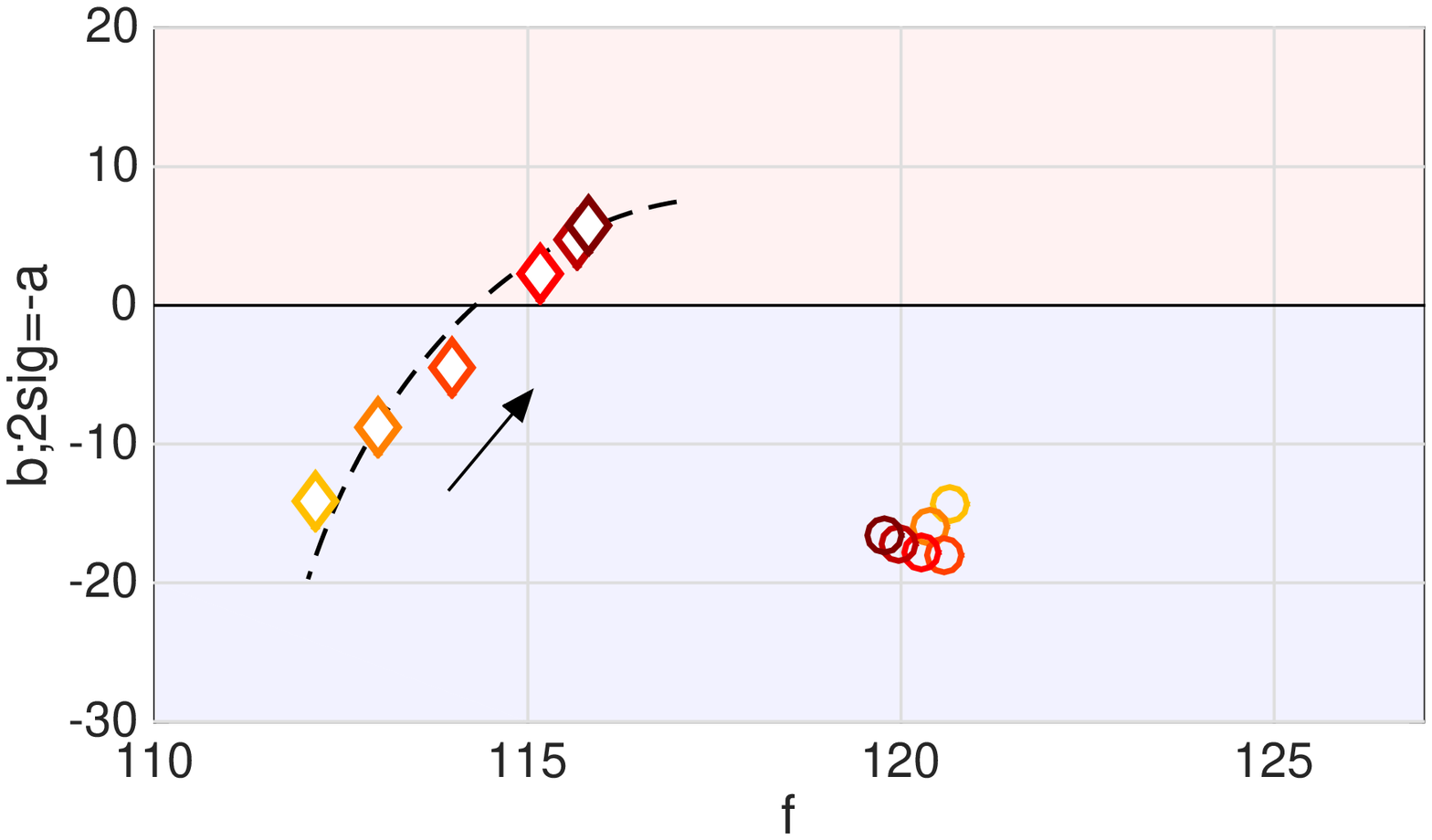}    
      \put(-7,57){$(b)$}
      \put(36, 25){\small \textcolor{black}{$\Phi$}}
      \put(27, 49){\small \textcolor{black}{$\pole_{ta}=(f_{ta},\nu)$}}
      \put(58, 15){\small \textcolor{black}{$\pole_{a}=(f_a,-\dfrac{\alpha}{2})$}}  
      \put(80, 41){\textcolor{red}{unstable}}
      \put(85, 33){\textcolor{blue}{stable}}       
   \end{overpic} 
}
\caption{
$(a)$  
Identified acoustic damping $\alpha$ (mean value 37~rad/s), heat release source strength $\beta$ and thermoacoustic growth rate $\nu$ vs. equivalence ratio.
$(b)$ 
Evolution of the system with equivalence ratio $\Phi = 0.521, \ldots, 0.549$ in a frequency-growth rate plane.
The acoustic pole $\pole_a$ is fixed (circles) whereas the  thermoacoustic pole $\pole_{ta}$  moves along a curve as $\Phi$ increases (diamonds),
which results in the system becoming unstable for $\Phi \simeq 0.536$, and in a frequency shift.
A simple model of time delay (dashed curve) captures this behaviour.
}
\label{fig:poles_vs_n_and_phi}
\end{figure}

The effect of the growth rate on the potential $\mathcal{V}(A)$ and on the PDF $P(A)$, discussed in section~\ref{sec:ID1} and fig.~\ref{fig:V_P}$(a)$, can now be decomposed in terms of acoustic damping and  flame gain.
When $\alpha>\beta$ the stabilising term $\alpha A^2/4$ dominates and tends to keep the potential well close to small amplitudes;
when  $\alpha<\beta$ the destabilising term $-\beta A^2/4$ dominates and tends to push the potential well towards larger amplitudes.

Figure~\ref{fig:poles_vs_n_and_phi}$(b)$ shows
 in a  frequency--growth rate plane the evolution with equivalence ratio $\Phi$ of the poles of the system:
 identified acoustic pole $\pole_a$  (circles)
and  identified thermoacoustic pole $\pole_{ta}$ (diamonds).
Their respective growth rates $-\alpha/2$ and $\nu=(\beta-\alpha)/2$ directly come from the system identification. 
The increase in thermoacoustic growth rate illustrates the influence of the flame  (specifically, the linear term $\beta \eta$ in the coherent heat release rate) on the system's stability.
The thermoacoustic frequency is approximated by  the measured frequency, $f_{ta} \simeq f_p$, since growth rates are small compared to pulsations. 
While the acoustic pole is independent of the equivalence ratio, the thermoacoustic pole moves along a curve as $\Phi$ increases, toward the unstable region and toward larger frequencies.

The frequency drift can be explained by considering that the time delay and gain from acoustic fluctuations $\eta(t)$ 
to heat release rate fluctuations $q_c(t)$ 
at the frequency of interest $\omega_0$ depend on the equivalence ratio.
 Using the simple time delay description 
 $\gamma \hat q_c = G(\tau) e^{-i \omega_{0} \tau} \hat \eta$ 
(equivalent to keeping in (\ref{eq:Taylor}) the coefficient $a_{0,1}$ in addition to $a_{1,0}$), 
the thermoacoustic growth rate vary as
$\nu = (G \cos(\omega_{0} \tau) -\alpha)/2$
and the pulsation as 
$\omega_{ta}^2 = \omega_{0}^2 - G \omega_{0}  \sin(\omega_{0} \tau)$.
Assuming linear variations for  $\tau(\Phi)$ and
$G(\Phi)$, one can fit the data and   satisfactorily retrieve the  simultaneous increase in growth rate and frequency, as shown by the dashed curved in figure~\ref{fig:poles_vs_n_and_phi}$(b)$.
This simple description 
could be refined by considering a  distribution of time delays over the spatial extent the flame \cite{Sattelmayer02, Schuermans03} or the superposition of two separate delay mechanisms via the axial and azimuthal convection of perturbations \cite{Palies11,Bade13}, as well as nonlinear frequency drifts resulting from amplitude-dependent time delays \cite{Noiray08};
therefore, we do not attempt to estimate quantitative time delay values.
Despite its simplicity, this description is consistent with measurements and accounts for the instability mechanisms at play in the combustor considered in this study, both in terms of frequency and growth rate.

\section{Conclusion}

\begin{table}
  \begin{center}
\def~{\hphantom{0}}
  \begin{tabular}{ll}
  Identification of  growth rate $\nu$  
& Identification of  acoustic damping $\alpha$ \\
\hline 
  1. Measure acoustic pressure $p(t)$;   
& 1. Measure acoustic pressure $p(t)$  
\\ 
  2. Band-pass filter around peak frequency $f_p \quad$ 
& $\quad$ and heat release rate $q(t)$; 
\\
$\quad\rightarrow$ modal amplitude $\eta(t)$;  
& 2. Compute transfer function $H = \hat p/\hat q $; 
\\
  3. Hilbert transform $\rightarrow$ amplitude $A(t)$;   
& 3. Fit a model transfer function $\widetilde H(s)$; 
\\
  4. Compute coefficients of the FP equation; 
& 4. Extract dominant acoustic pole  
\\
  5. Fit analytical expressions $\rightarrow$ identify $\nu$.  
& $\quad$ $\sigma+i\omega \simeq -\alpha/2+i \omega_a$ $\rightarrow$ identify $\alpha$. 
  \end{tabular}
  \caption{Summary of the two system identification methods.}
  \label{tab:methods}
  \end{center}
\end{table}

A new system identification technique based on the processing of simultaneously-recorded acoustic pressure and flame chemiluminescence signals has been proposed in this paper. 
It constitutes a precious complement to the output-only SI approach proposed by \cite{NoiraySchu13}, which gives access to the linear growth rate of an observed thermoacoustic limit-cycle from the computation of the drift and diffusion coefficients of the Fokker-Planck equation describing the acoustic pressure statistics. 
These two methods are summarised in table~\ref{tab:methods}.
Together, they yield not only the linear growth rate, but also the modal acoustic damping and the linear contribution of the acoustic-flame coupling. 
It is important to note that the acoustic damping cannot  be deduced from the quality factor of thermoacoustic peaks associated with linearly stable operating conditions, because the effect of the flame is embedded into the observed dynamics. 
Thanks to this novel technique, the knowledge of the modal damping coefficients of a combustor 
will be particularly useful for validating low-order \emph{predictive} thermoacoustic network models. Therefore it constitutes a  substantial progress in the development of model-based SI methods for thermoacoustic instabilities.
The same principle can be applied to other types of instabilities, e.g. in aeroacoustic systems, provided source and damping terms are available.

\section*{Acknowledgements}

E. B. and N. N. acknowledge support by Repower and the ETH Zurich Foundation.

\bibliographystyle{plain}
\bibliography{chemi}

\clearpage
\newpage

\begin{center}
 \begin{LARGE}
  \textbf{Supplementary material}
 \end{LARGE}
\end{center}
\vspace{1cm}

\begin{center}
\psfrag{f}[t][]{$f$~(Hz)}
\psfrag{fff}[t][]{$f$~(Hz)}
\psfrag{ffff}[t][]{}
\psfrag{magnitude}[r][][1][-90]{$|H|$}
\psfrag{growthrate}[b][]{$\sigma_k, \Re(z_l)$ (rad/s)}
\psfrag{phase}[][][1][-90]{$\,\,\angle H$}
\hspace{0.2cm}
    \begin{overpic}[width=8.18cm,tics=10]{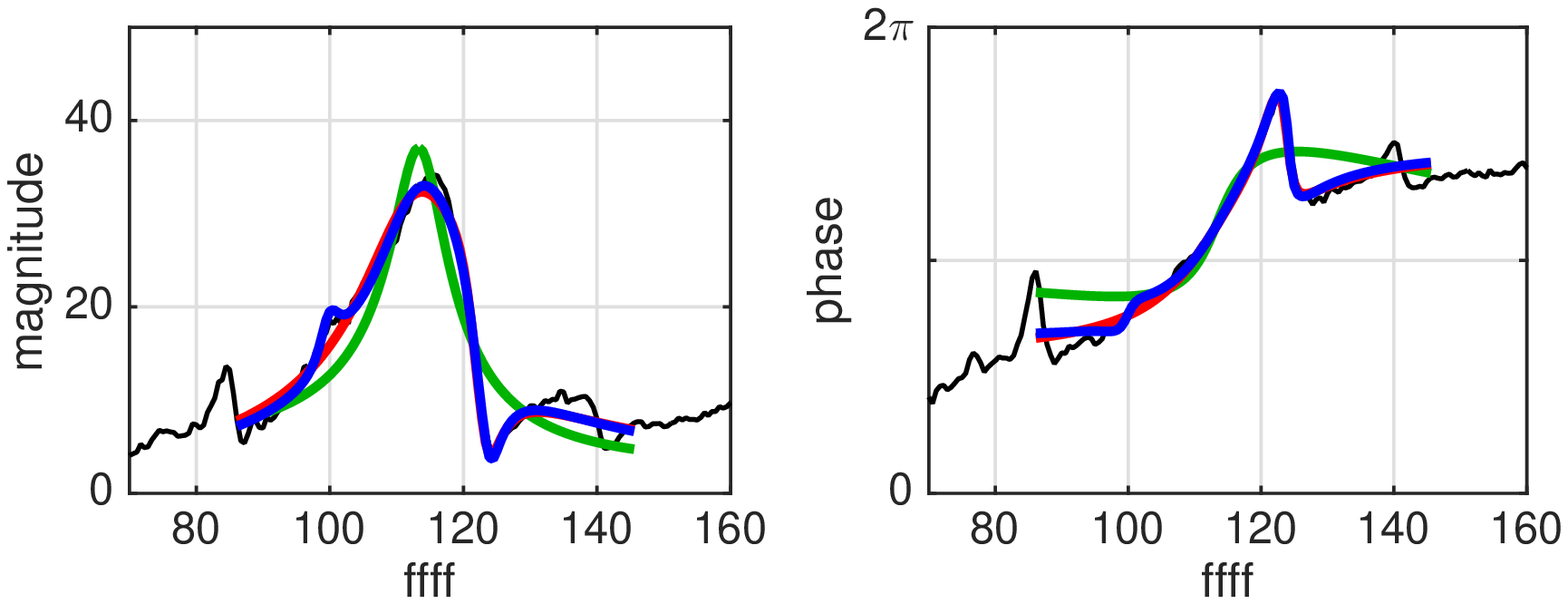}
      \put(-5,35){$(a)$}
      \put(74,9.5){$\Delta f=30$ Hz}
      \put(75,14){$\nTF=\textcolor[rgb]{0,0.7,0}{2}$, \textcolor[rgb]{1,0,0}{4}, \textcolor[rgb]{0,0,1}{6}}
   \end{overpic} 
   \hspace{0.5cm}
   \begin{overpic}[width=4.09cm,tics=10]{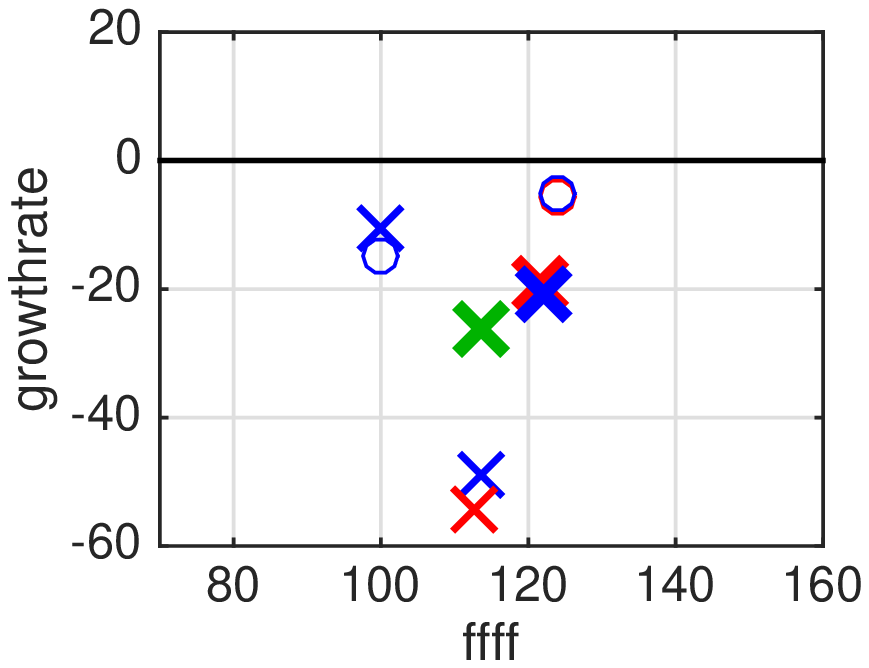}
   \end{overpic}  
\end{center}

\vspace{-0.15cm}

\begin{center}
\psfrag{f}[t][]{$f$~(Hz)}
\psfrag{fff}[t][]{$f$~(Hz)}
\psfrag{ffff}[t][]{}
\psfrag{magnitude}[r][][1][-90]{$|H|$}
\psfrag{growthrate}[b][]{$\sigma_k, \Re(z_l)$ (rad/s)}
\psfrag{phase}[][][1][-90]{$\,\,\angle H$}
\hspace{0.2cm}
   \begin{overpic}[width=8.18cm,tics=10]{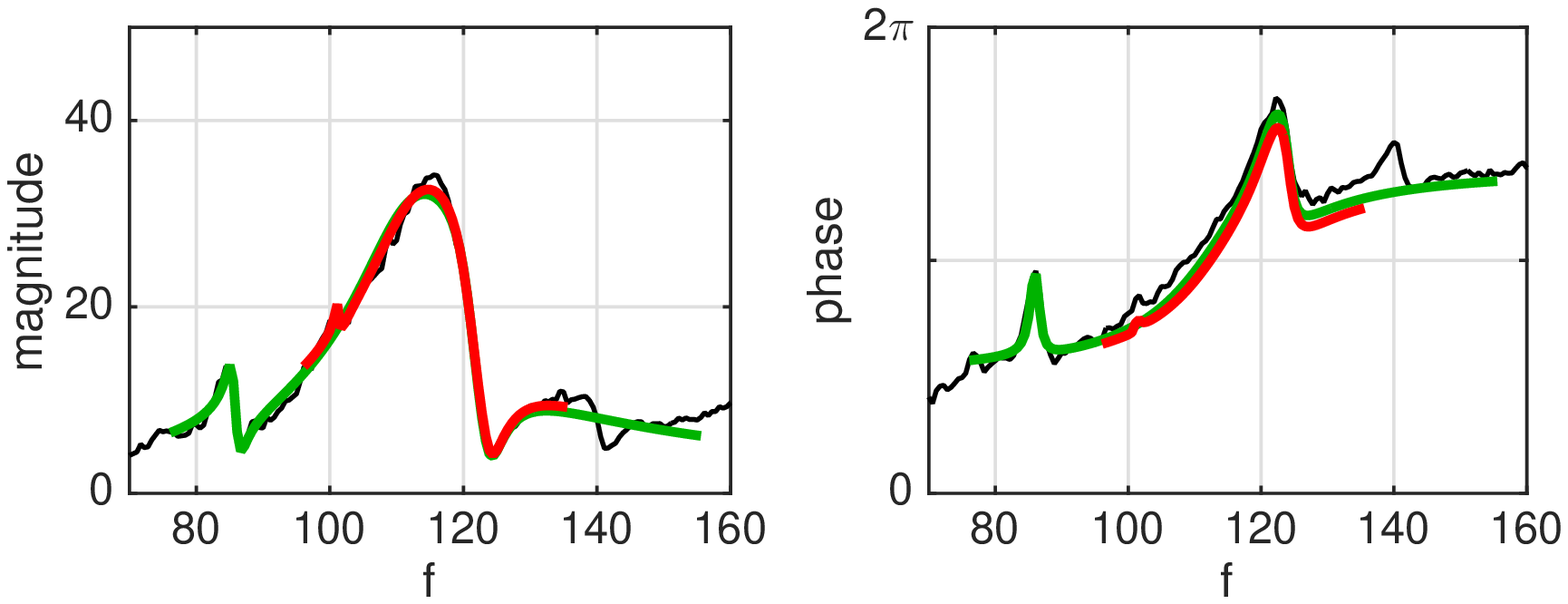}
      \put(-5,35){$(b)$}
      \put(67.5,9.5){$\Delta f=\textcolor{red}{20}$, $\textcolor[rgb]{0,0.7,0}{40}$ Hz}
      \put(83.5,14){$\nTF=6$}
   \end{overpic} 
   \hspace{0.5cm}
   \begin{overpic}[width=4.09cm,tics=10]{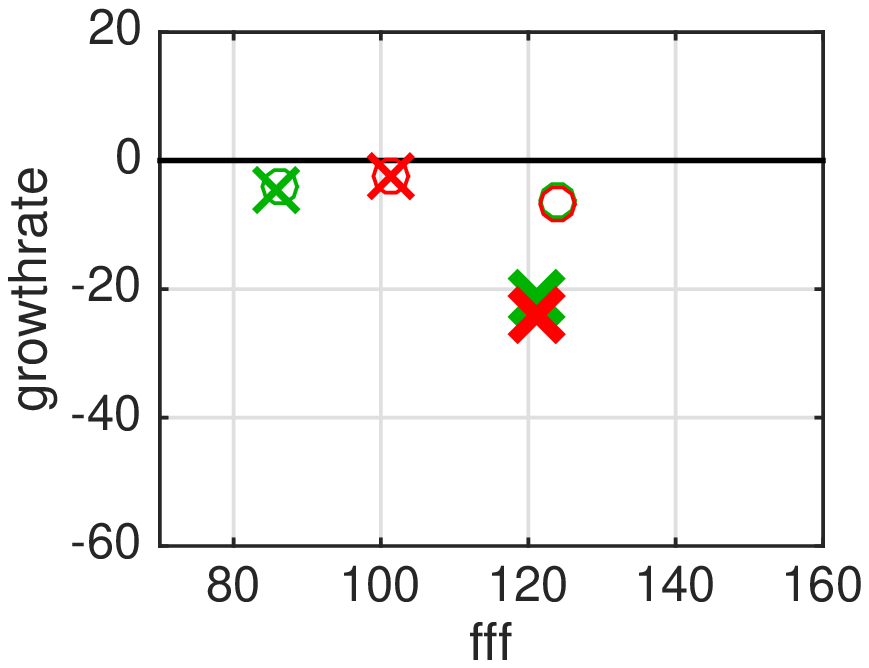}
   \end{overpic}
\captionof{figure}{Simultaneous fit of the gain and phase of the acoustic transfer function $H(f)$, and corresponding poles ($\times$) and zeros ({\Large$\boldsymbol{\cdot}$}) for different orders $\nTF$ and frequency ranges $f_p\pm\Delta f$:
$(a)$~$\nTF=2,$ 4 and 6, $\Delta f=30$~Hz;
$(b)$~$\nTF=6$, $\Delta f=20$ and 40~Hz.
Bold crosses indicate the dominant pole $\pole_a$ in the vicinity of the peak frequency $f_p$.
Equivalence ratio $\Phi=0.538$.}
\label{fig:fit}
\end{center}

\vspace{4cm}

\begin{center}
\psfrag{phi}[t][]{$\Phi$~(-)}
\psfrag{n}[t][]{$\nTF$}
\psfrag{growthrate}[][]{$\sigma$~(rad/s)}
\psfrag{gg}[r][][1][-90]{}
\psfrag{freq}[r][][1][-90]{$f$~(Hz)}
\psfrag{sigma}[][]{$\sigma$~(rad/s)}
\psfrag{alpha}[r][][1][-90]{$\alpha$~(rad/s)}
\psfrag{beta}[r][][1][-90]{$\beta$~(rad/s)}
\psfrag{ab}[][]{$\alpha, \beta$~(rad/s)}
\psfrag{2nu}[][]{$2\nu$~(rad/s)}
   \begin{overpic}[width=4.1cm,tics=10]{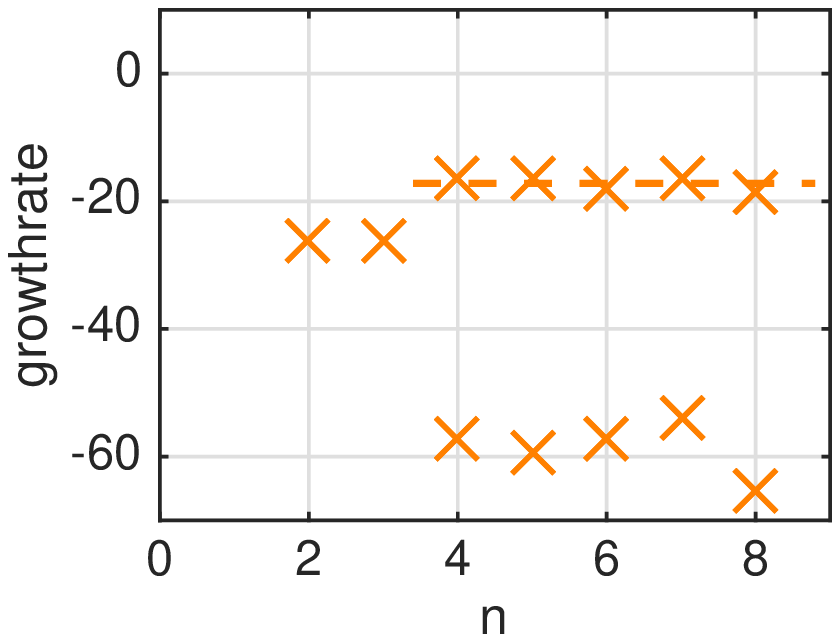}
      \put(42,81){$\Phi=0.526$}
      \put(55,62){$\sigid=-\alpha/2$}
      \put(20,81){$(a)$}
   \end{overpic} 
  \hspace{0.3cm}  
   \begin{overpic}[width=4.1cm,tics=10]{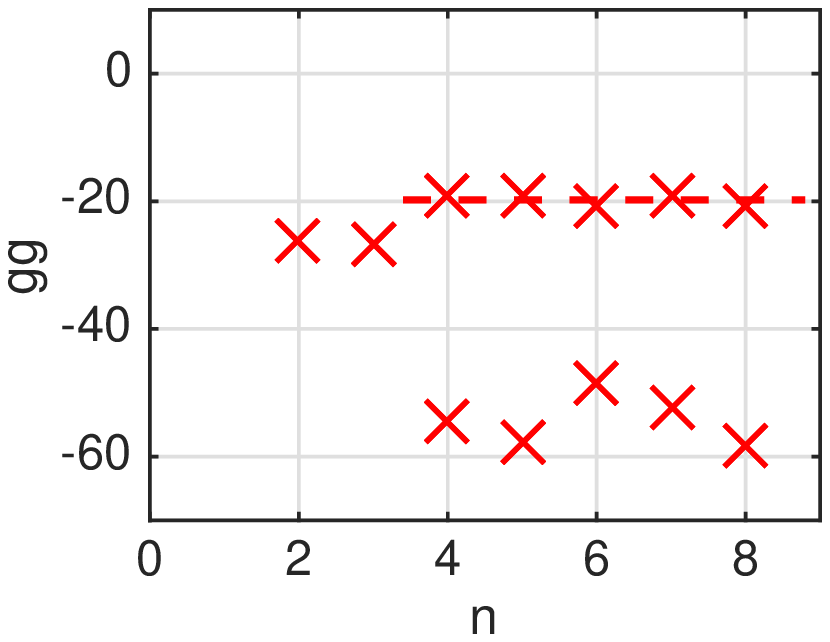}
      \put(42,81){$\Phi=0.538$}
      \put(55,60){$\sigid=-\alpha/2$}
      \put(20,81){$(b)$}
   \end{overpic} 
   \hspace{0.3cm}   
   \begin{overpic}[width=4.1cm,tics=10]{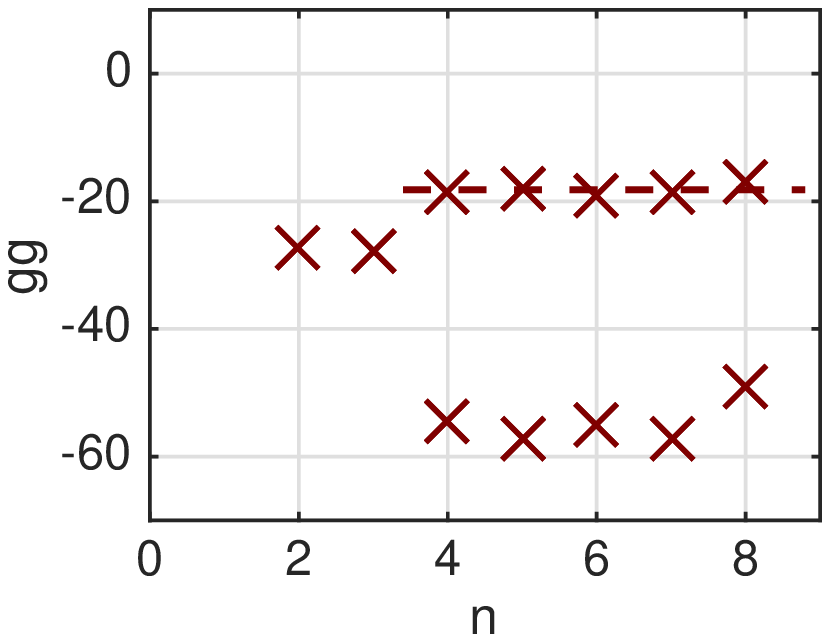}                                        
      \put(42,81){$\Phi=0.549$}
      \put(55,61){$\sigid=-\alpha/2$}
      \put(20,81){$(c)$}
   \end{overpic}    
\captionof{figure}{Real part of the  poles identified in the vicinity of the peak frequency $f_p$, vs. order $\nTF$ of the fitting transfer function. 
  $(a)$ $\Phi=0.526$,  $(b)$  $\Phi=0.538$,   $(c)$ $\Phi=0.549$. Dashed lines show the real part $\sigid$ of the identified dominant pole.
}
\label{fig:poles_vs_n_and_phi}
\end{center}

\end{document}